\documentclass[%
 preprint,
 amsmath,amssymb,
 aps,
]{revtex4-1}

\usepackage{graphicx}% Include figure files
\usepackage{subfigure}% Include figure files
\usepackage{dcolumn}% Align table columns on decimal point
\usepackage{bm}% bold math
\usepackage[flushleft]{threeparttable}
\usepackage{url}
\usepackage{amsmath}
\usepackage{amssymb}
\usepackage{braket}
\usepackage{multirow}
\usepackage{array}
\usepackage{url}
\usepackage{booktabs}
\usepackage{xcolor}
\usepackage{framed}
\usepackage{lipsum}
\colorlet{shadecolor}{yellow!30}
\newcolumntype{P}[1]{>{\centering\arraybackslash}p{#1}}

\begin{document}

\preprint{APS/123-QED}
\title{Atomistic Description for Temperature--Driven Phase Transitions in BaTiO$_3$}

\author{Y. Qi,$^1$ S. Liu,$^{1,2}$ I. Grinberg,$^1$ and A. M. Rappe$^1$}

\affiliation{%
 $^1$The Makineni Theoretical Laboratories, Department of Chemistry,\\
 University of Pennsylvania, Philadelphia, PA 19104-6323, United States\\
 $^2$Geophysical Laboratory Carnegie Institution for Science, Washington, D.C. 20015, United States
}%

\begin{abstract}
Barium titanate (BaTiO$_3$) is a prototypical ferroelectric perovskite that undergoes the rhombohedral--orthorhombic--tetragonal--cubic phase transitions as the temperature increases. 
In this work, we develop a classical interatomic potential for BaTiO$_3$ within the framework of the bond--valence theory.
The force field is parameterized from first--principles results, enabling accurate large--scale molecular dynamics (MD) simulations at finite temperatures. 
Our model potential for BaTiO$_3$ reproduces the temperature--driven phase transitions in isobaric--isothermal ensemble ($NPT$) MD simulations. 
This potential allows the analysis of BaTiO$_3$ structures with atomic resolution.
By analyzing the local displacements of Ti atoms, we demonstrate that the phase transitions of BaTiO$_3$ exhibit a mix of order--disorder and displacive characters.
Besides, from detailed observation of structural dynamics during phase transition, 
we discover that the global phase transition is associated with changes in the equilibrium value and fluctuations of each polarization component, including the ones already averaging to zero, 
Contrary to the conventional understanding that temperature increase generally causes bond--softening transition, 
the $x$ polarization component exhibits a bond--hardening character during the orthorhombic to tetragonal transition.
These results provide further insights about the temperature--driven phase transitions in BaTiO$_3$.

\end{abstract}

\pacs{Valid PACS appear here}% PACS, the Physics and Astronomy
\maketitle
\section{Introduction}

BaTiO$_3$ is a ferroelectric perovskite with promising applications in electronic devices, 
such as non--volatile memory, high--$\kappa$ dielectrics, and piezoelectric sensors~\cite{Yun02p447,Chen07p553,Buessem66p33,Karaki07p7035}.
Therefore, it is of great significance to investigate and understand the structural and electronic properties of BaTiO$_3$ for designed material optimization and device engineering. 
First--principles density functional theory (DFT) has served as a powerful method to understand the electronic structures of ferroelectric materials~\cite{Cohen92p65,Gonze97p10355,Saha00p8828,Kolpak08p036102,Martirez12p256802,Morales14p4465}.
Due to the expensive computational cost, the application of DFT methods is currently limited to system of fairly small size at zero Kelvin. 
Many important dynamical properties, such as domain wall motions and temperature--driven phase transitions, are beyond the capability of conventional first--principles methods. 
An effective Hamiltonian method was developed to study finite--temperature properties of BaTiO$_3$~\cite{Zhong94p1861,Zhong95p6301,Nishimatsu08p104104,Fu03p257601}.
To apply this method, the subset of dynamical modes that determine a specific property should be known {\em a priori}. 
Molecular dynamics (MD) simulations with an atomistic potential accounting for all the modes offer distinct advantages,
especially in providing detailed information about atomic positions, velocities and modifications of chemical bonds due to a chemical reaction or thermal excitation. 
The shell model for BaTiO$_3$ has been developed~\cite{Tinte99p9679,Tinte01p235403,Tinte00p41,Zhang10p015701,Zhang09p405703}. 
However, due to the low mass assigned to the shell, a small time step in MD simulations is required to achieve accurate results, which limits the time and length scales of the simulations. \\

Recently, we developed a bond--valence (BV) model potential for oxides based on the bond valence theory~\cite{Grinberg02p909,Shin08p015224,Shin05p054104,Liu13p102202,Liu13p104102}.
The force fields for many technologically important ferroelectric materials, PbTiO$_3$, PbZrO$_3$ and BiFeO$_3$~\cite{Liu13p102202,Liu13p104102,Shin05p054104,Grinberg02p909,Chen15p6371},
have been parameterized based on results from DFT calculations. 
A typical force field requires no more than 15 parameters and can be efficiently implemented, 
which enables simulations of systems with thousands of atoms under periodic boundary conditions~\cite{Liu13p232907,Xu15p79}. 
The development of an accurate classical potential for BaTiO$_3$ has proven to be difficult, 
mainly due to the small energy differences among the four phases (rhombohedral, orthorhombic, tetragonal, and cubic)~\cite{Kwei93p2368,Dieguez04p212101,Von50p221}. 
In this paper, we apply the bond--valence model to BaTiO$_3$ and parameterize the all--atom interatomic potential to first--principles data. 
Our model potential for BaTiO$_3$ is able to reproduce the rhombohedral--orthorhombic--tetragonal--cubic (R-O-T-C) phase transition sequence in isobaric--isothermal ensemble ($NPT$) MD simulations. 
The phase transition temperatures agree reasonably well with previous theoretical results~\cite{Tinte99p9679}.  
We further examine the temperature dependence of the local displacements of Ti atoms and discover several features of the phase transitions of BaTiO$_3$:
the phase transitions of BaTiO$_3$ involve both order--disorder and displacive characters;
at the moment that the phase transition of the crystal occurs, all the polarization components undergo phase transitions, even for the nonpolar ones;
and temperature increase can also cause bond--hardening for a certain component. 

\section{Methods}

\begin{figure}[htbp]
\includegraphics[width=12.0cm]{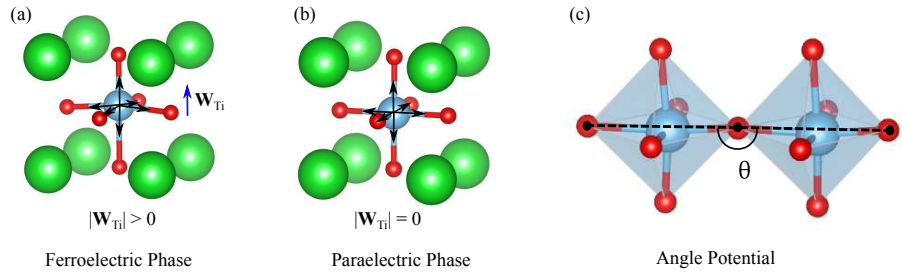}
\caption{Bond valence vector sum and angle potential. (a) Tetragonal BaTiO$_3$ with a non-zero BVVS; (b) Cubic BaTiO$_3$ with zero BVVS;
(c) Schematic of the angle potential. Ba, Ti, and O atoms are represented by green, blue and red spheres respectively.
}\label{f1}
\end{figure}

The bond--valence model potential is developed based on the conservation principles of bond valence and bond--valence vector. 
The bond valence, $V_{ij}$, reflects the bonding strength and can be calculated based on the bond length, $r_{ij}$, with~\cite{Brown09p6858,Grinberg02p909,Brown73p266,Brown76p1957,Liu13p102202,Liu13p104102,Shin08p015224,Shin05p054104}
\begin{equation}
V_{ij}=\left(\frac{r_{0,ij}}{r_{ij}}\right)^{C_{ij}}
\end{equation}
where $i$ and $j$ are the labels for atoms; $r_{0,ij}$ and $C_{ij}$ are Brown's empirical parameters. The bond--valence vector is defined as a vector lying along the bond,
${\bf{V}}_{ij}=V_{ij}\hat{{\bf{R}}}_{ij}$, where $\hat{{\bf{R}}}_{ij}$ is the unit vector pointing from atom $i$ to atom $j$. 
The total energy ($E$) consists of the Coulombic energy ($E_c$), 
the short--range repulsive energy ($E_r$), the bond--valence energy ($E_{\rm BV}$), the bond--valence vector energy ($E_{\rm BVV}$), and the angle potential ($E_a$)~\cite{Liu13p102202,Liu13p104102,Shin08p015224,Shin05p054104}: 
\begin{equation}
E=E_c+E_r+E_{\rm BV}+E_{\rm BVV}+E_a
\end{equation}
\begin{equation}
E_c=\sum_{i<j}\frac{q_iq_j}{r_{ij}},
\end{equation}
\begin{equation}
E_r=\sum_{i<j}\left(\frac{B_{ij}}{r_{ij}}\right)^{12},
\end{equation}
\begin{equation}
E_{\rm BV}=\sum_{i}S_i\left(V_i-V_{0,i}\right)^2
\end{equation}
\begin{equation}
E_{\rm BVV}=\sum_iD_i\left({\bf{W}}_i^2-{\bf{W}}_{0,i}^2\right)^2
\end{equation}
\begin{equation}
E_{a}=k\sum_i^{N_{\rm oxygen}}\left(\theta_i-180^\circ\right)^2
\end{equation}
where ${V}_i=\sum_{j\not=i}V_{ij}$ is the bond--valence sum (BVS), ${\bf{W}}_i=\sum_{j\not=i}{\bf{V}}_{ij}$ is the bond--valence vector sum (BVVS, shown in FIG.~\ref{f1} (a), (b)),
$q_i$ is the ionic charge, $B_{ij}$ is the short--range repulsion parameter, $S_{i}$ and $D_i$ are scaling parameters with the unit of energy,
$k$ is the spring constant and $\theta$ is the O--O--O angle along the common axis of two adjacent oxygen octahedra (FIG.~\ref{f1} (c)).
The bond-valence energy $E_{\rm BV}$ captures the energy penalty for both overbonded and underbonded atoms. 
The bond-valence vector energy $E_{\rm BVV}$ is a measure of the breaking of local symmetry, which is important for correctly describing the ferroelectricity. 
$V_{0,i}$ and ${\bf{W}}_{0,i}$ are preferred or target values of BVS and BVVS for atom $i$ in the ground--state structure, 
which can be calculated from DFT directly. It is noted that the $E_{\rm BV}$  and $E_{\rm BVV}$ can be related to the moments of the local density of states in the framework of a tight binding model, 
providing a quantum mechanical justification for these two energy terms~\cite{Liu13p102202,Liu13p104102,Finnis84p45,Brown09p6858,Harvey06p1038}. 
The angle potential is used to account for the energy cost associated with the rotations of oxygen octahedra.  \\

We followed the optimization protocol developed in previous studies~\cite{Liu13p102202,Liu13p104102}.
The optimal values of force-field parameters $q_i$, $S_i$, $D_i$ and $B_{ij}$, are acquired by minimizing the difference between the DFT energies/forces and the model--potential energies/forces for a database of BaTiO$_3$ structures.
All DFT calculations are carried out with the plane--wave DFT package \textsc{Quantum--espresso}~\cite{Giannozzi09p395502etal} 
using the Perdew--Burke--Ernzerhof functional modified for solids (PBEsol)~\cite{Perdew08p136406} and optimized norm--conserving pseudopotentials generated by the \textsc{Opium} package~\cite{Opium}.
A plane--wave cutoff energy of 50 Ry and 4$\times$4$\times4$ Monkhorst--Pack $k$--point mesh~\cite{Monkhorst76p5188} are used for energy and force calcualtions. 
The database consists of 40--atom 2$\times$2$\times$2 supercells with different lattice constants and local ion displacements. The final average difference between DFT energy and model--potential energy is 1.35 meV/atom.

\section{Performance of the classical potential}

\begin{figure}[htbp]
\includegraphics[width=8.0cm]{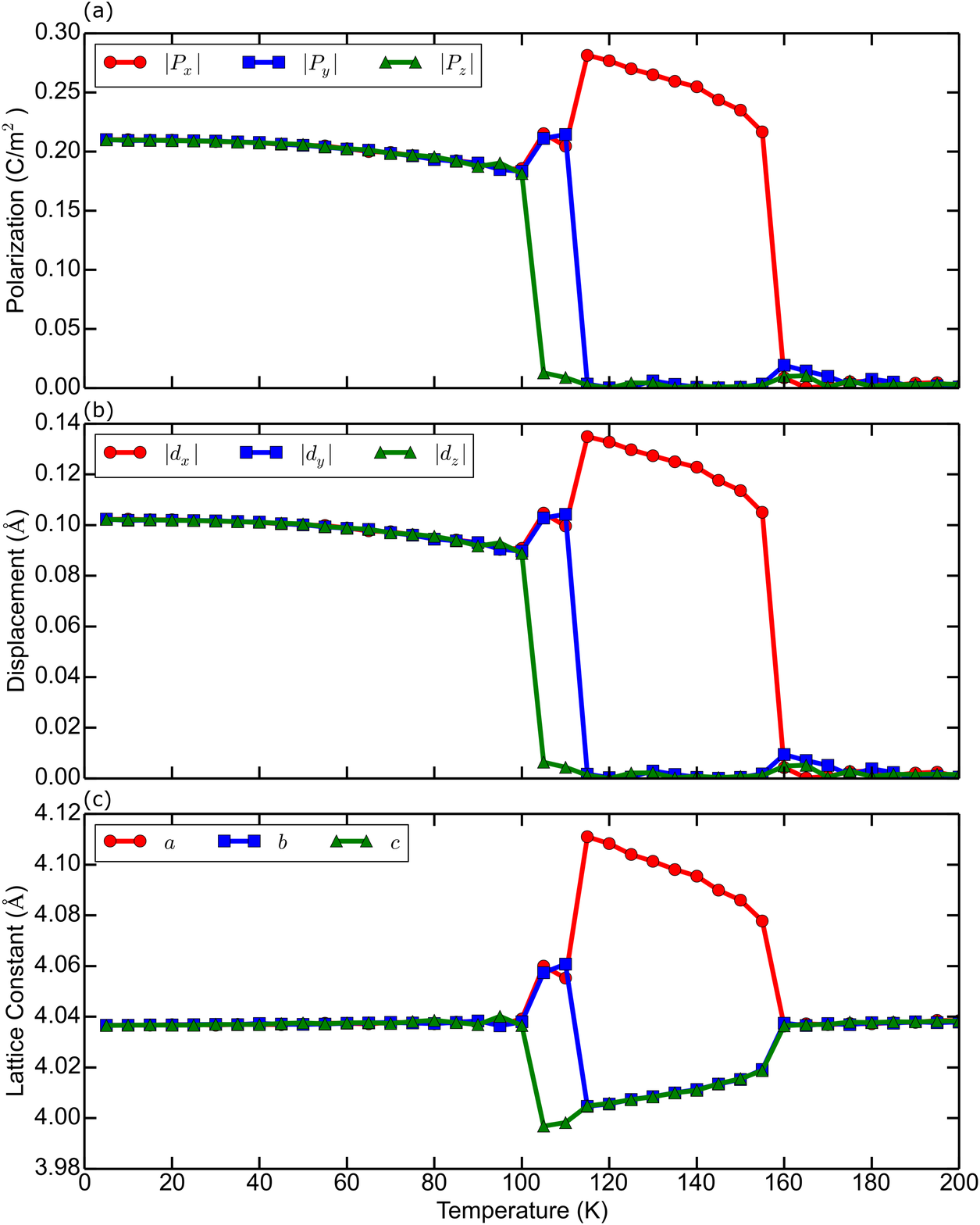}
\caption{Temperature dependence of the polarization, Ti displacement, and lattice constants in BaTiO$_3$. 
Phase transitions between rhombohedral, orthorhombic, tetragonal, and cubic occur at 105 K, 115 K and 160 K.}\label{f2}
\end{figure}

The optimized parameters are listed in TABLE~\ref{t1}. 
The performance of the obtained force field is examined by investigating the temperature dependence of lattice constants ($a$, $b$ and $c$), 
component--resolved local displacements of Ti atoms ($d_x$, $d_y$, and $d_z$), and the three components of the total polarization ($P_x$, $P_y$, and $P_z$).
We carry out $NPT$ MD simulations using a 10$\times$10$\times$10 supercell (5000 atoms) with the temperature controlled via the Nos{\'e}--Hoover thermostat
and the pressure maintained at 1 atm via the Parrinello--Rahman barostat~\cite{Parrinello80p1196}.
As shown in FIG.~\ref{f2}, the simulations clearly reveal four distinct phases under different temperature ranges and three first--order phase transitions. 
Below 100~K, the displacements of Ti atoms and the overall polarization of the supercell are along [111] direction ($P_x=P_y=P_z$), characteristic of the rhombohedral phase. 
At 100~K, the $z$ component of the total polarization, $P_z$, becomes approximately 0, indicating a phase transition from rhombohedral to orthorhombic ($P_x=P_y>0$, $P_z=0$).
As the temperature increases further to 110 K, the total polarization aligns preferentially along $x$ direction ($P_x>0$, $P_y=P_z=0$) and the lattice constants have $b=c<a$. 
The supercell stays tetragonal until 160 K at which point the ferroelectric--paraelectric phase transition occurs. 
The phase transition temperatures match well with those predicted by the shell model~\cite{Tinte99p9679} (TABLE~\ref{t2}).
The underestimation of Curie temperature $T_C$ in MD simulations has been observed previously and is likely due to the systematic error of density functional used for force field optimization~\cite{Liu13p102202,Sepliarsky05p014110}. 
We extract the averaged lattice constants at finite temperatures from MD simulations and find that they are in good agreement (error less than 1$\%$) with the PBEsol values (TABLE~\ref{t3}). 

Domain walls are interfaces separating domains with different polarities. They are important topological defects and can be moved by applying external stimulus~\cite{Liu13p232907,Xu15p79}. 
The domain wall energy for a 180$^{\circ}$ wall obtained from our MD simulations is 6.63 mJ/m${^2}$, 
which is comparable to PBEsol value, 7.84 mJ/m$^2$. This indicates that our atomistic potential can be used for studying the dynamics of ferroelectric domain walls in BaTiO$_3$.
All these results demonstrate the robustness of this developed classical potential.

\section{Atomistic features of different phases}

To provide an atomistic description of the different phases of BaTiO$_3$, 
we analyze the distribution of local displacements of Ti atoms in each phase.
Ti displacement is defined as the distance between the Ti atom and the center of the oxygen octahedral cage of a unit cell,
which scales with the magnitude of polarization.

\begin{figure}[htbp]
\includegraphics[width=12.0cm]{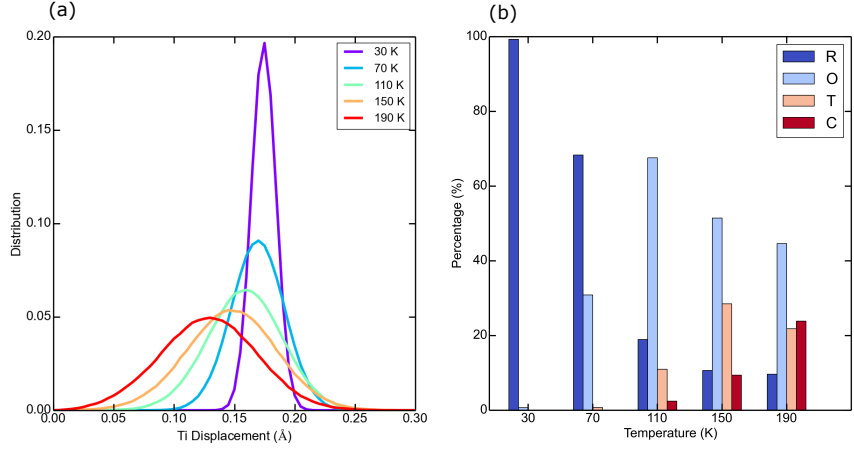}
\caption{(a) The distribution of total Ti displacement magnitude at different temperatures.
(b) Compositions of different phases. Supercells at 30 K (rhombohedral), 70 K (rhombohedral), 110 K (orthorhombic), 150 K (tetragonal) and 190 K (cubic) are studied. 
Heights of the dark blue, light blue, orange, and red rectangles represent the percentages of rhombohedral, orthorhombic, tetragonal and cubic unit cells respectively.
The phases of unit cells are categorized by their Ti displacements $d$:
for $d<0.1$ \AA, the unit cell is considered as a nonpolar one;
for a polar unit cell, if one component is larger than $d/\sqrt{6}$, this component is considered as a ferroelectric one. 
The ferroelectric phase (tetragonal, orthorhombic and rhombohedral) is determined by the number of ferroelectric components.
}\label{f3}
\end{figure}

In FIG.~\ref{f3} (a), we plot the distributions of Ti displacements ($d=\sqrt{d_x^2+d_y^2+d_z^2}$). 
It can be seen that in all four phases, the distribution is approximately a Gaussian curve whose peak shifts toward lower values as the temperature increases. 
This suggests that the temperature--driven phase transition has a displacive character. 
It is noted that the distribution of magnitudes is peaked at non--zero value even in the paraelectric cubic phase, 
suggesting that most Ti atoms are still locally displaced at high temperature, and that the overall net zero polarization is the result of an isotropic distribution of local dipoles along different directions.
This confirms the order--disorder character for BaTiO$_3$ at high temperature. 

We can categorize the phase of each unit cell based on the local displacement of Ti atom. The categorization criteria are\\
(1) If $d<0.1$ \AA, the unit cell is considered to be paraelectric cubic;\\
(2) For a ferroelectric unit cell, the $k$--th component is considered to be ferroelectric if $d_k > d/\sqrt{6}$. The rhombohedral, orthorhombic, and tetragonal unit cells have three, two, and one ferroelectric component(s), respectively. \\
\noindent The results are shown in FIG.~\ref{f3} (b). 
At 30 K, the supercell is made only from rhombohedral unit cells, showing that the rhombohedral phase is the ground--state structure. As the temperature increases, the supercell becomes a mixture of the four phases. 
It should be noted that the cubic unit cell with nearly--zero local Ti displacement seldom appears, because a cubic unit cell is energetically less favorable. 
The relative energies of the four phases of BaTiO$_3$ from PBEsol DFT calculations are listed in TABLE~\ref{t5}. 
It can be seen that the energy differences between the tetragonal, orthorhombic and rhombohedral unit cells are small (within several meV per unit cell)~\cite{Cohen90p6416,Cohen92p65}. 
Due to the thermal fluctuations, the populations of higher--energy ferroelectric phases (tetragonal and orthorhombic) increase as temperature increases.
Above the ferroelectric--paraelectric  transition temperature,  locally ferroelectric unit cells are still favored over paraelectric due to the relatively high energy of cubic, the high--symmetry structure.  

\begin{figure}[htbp]
\includegraphics[width=8.0cm]{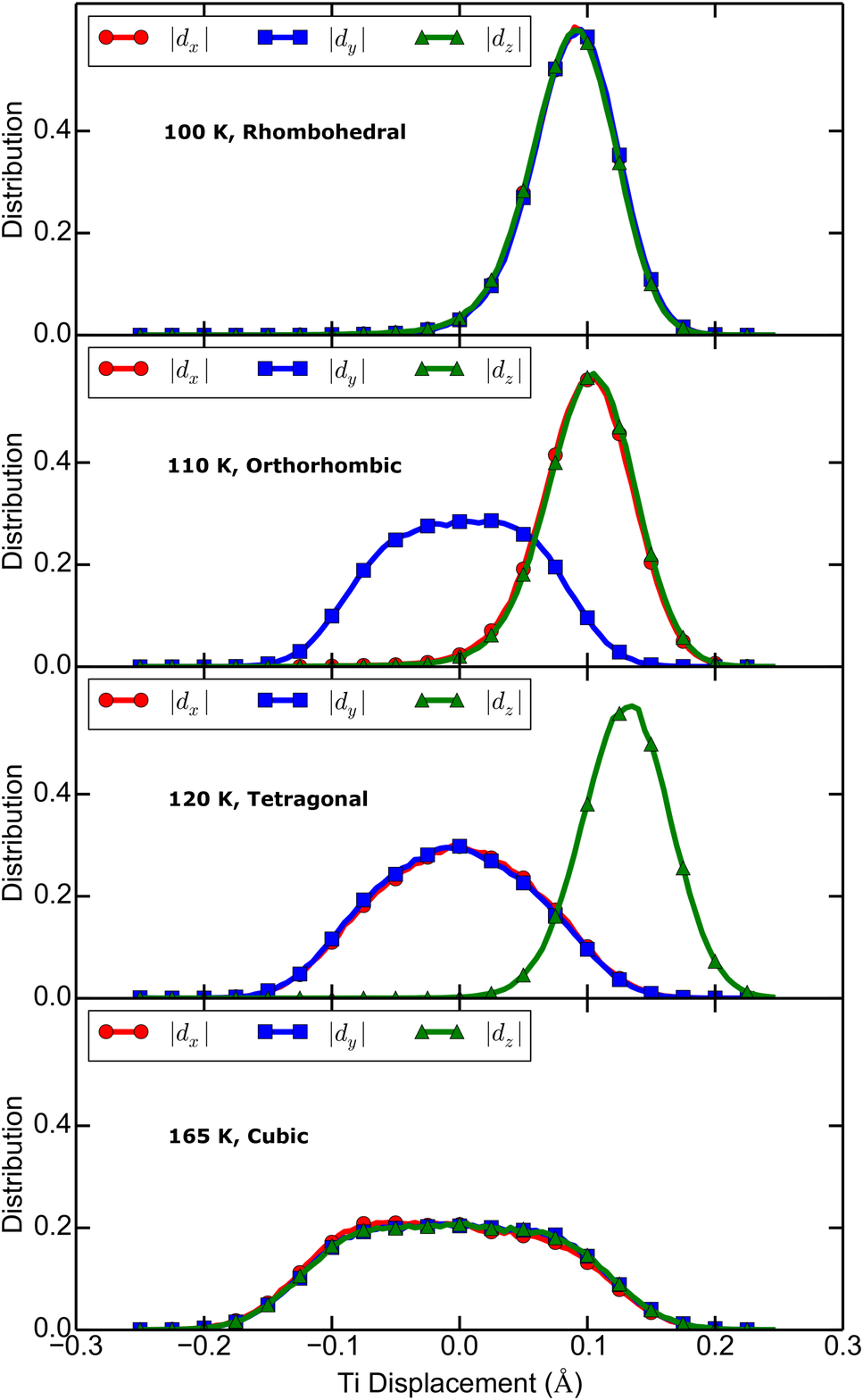}
\caption{The distributions of Ti displacement at different temperatures.}\label{f4}
\end{figure}

\begin{figure}[htbp]
\includegraphics[width=8.0cm]{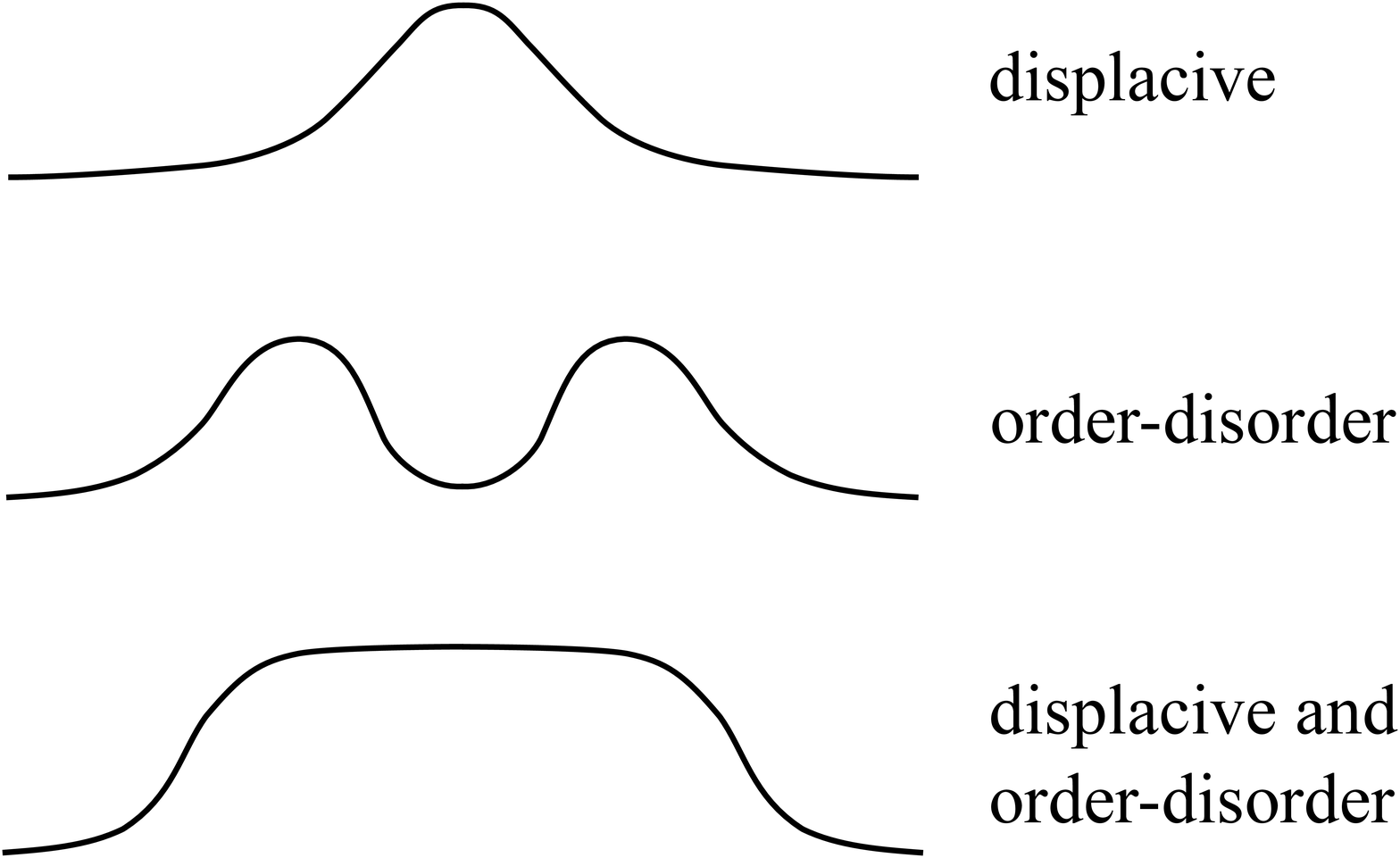}
\caption{Schematic figure of the distributions of Ti displacement for displacive transition, order--disorder transition and a mix of them.}\label{fadd}
\end{figure}

In FIG.~\ref{f4}, the distributions of Ti displacements along the three axes are plotted. 
At 100 K, BaTiO$_3$ is at the rhombohedral phase and the distributions of Ti displacements are Gaussian--like.
As the temperature increases, the phase changes to orthorhombic.
The average of the $x$ polarization component shifts to zero, indicating a displacive phase transition.
Besides, the standard deviation increases and the center of the distribution curve becomes flatter.
For the cubic phase, the center of the Ti displacement distribution curve is also flat.
As shown in FIG.~\ref{fadd}, the center--flat curve is a summation of a Gaussian curve centering at zero, 
and a double--peak curve. The latter is characteristic of order--disorder transition~\cite{Liu13p232907}.
These results further demonstrate that phase transitions of BaTiO$_3$ have a mix of order--disorder and displacive characters~\cite{Gaudon15p6,Ehses81p507,Jun88p2255,Comes68p715,Itoh85p29,Stern04p037601,Kwei93p2368}. 

\section{Features of the phase transitions}
\begin{figure}[htbp]
\includegraphics[width=15.0cm]{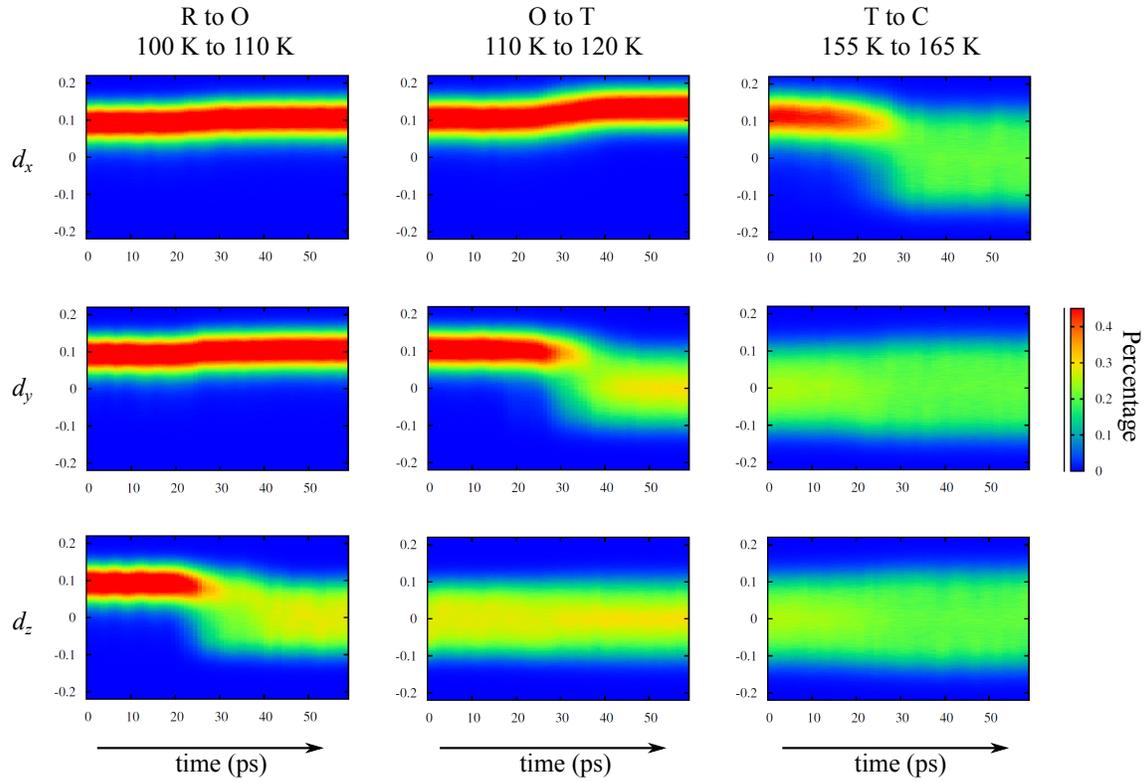}
\caption{Temperature dependence of Ti displacement distributions in three Cartesian directions. 
The horizontal axis shows the time. In these simulations, the temperature increases with time approximately linearly.
The vertical axis represents the fraction of the Ti displacements and the color scale represents the percentages of Ti displacement with a certain value.
Note that in the bottom center plot, the color showing the distribution becomes redder after the orthorhombic to tetragonal transition, 
indicating a narrower distribution around $d_z=0$ and a bond hardening in this direction.}\label{f5}
\end{figure}

\begin{figure}[htbp]
\includegraphics[width=10.0cm]{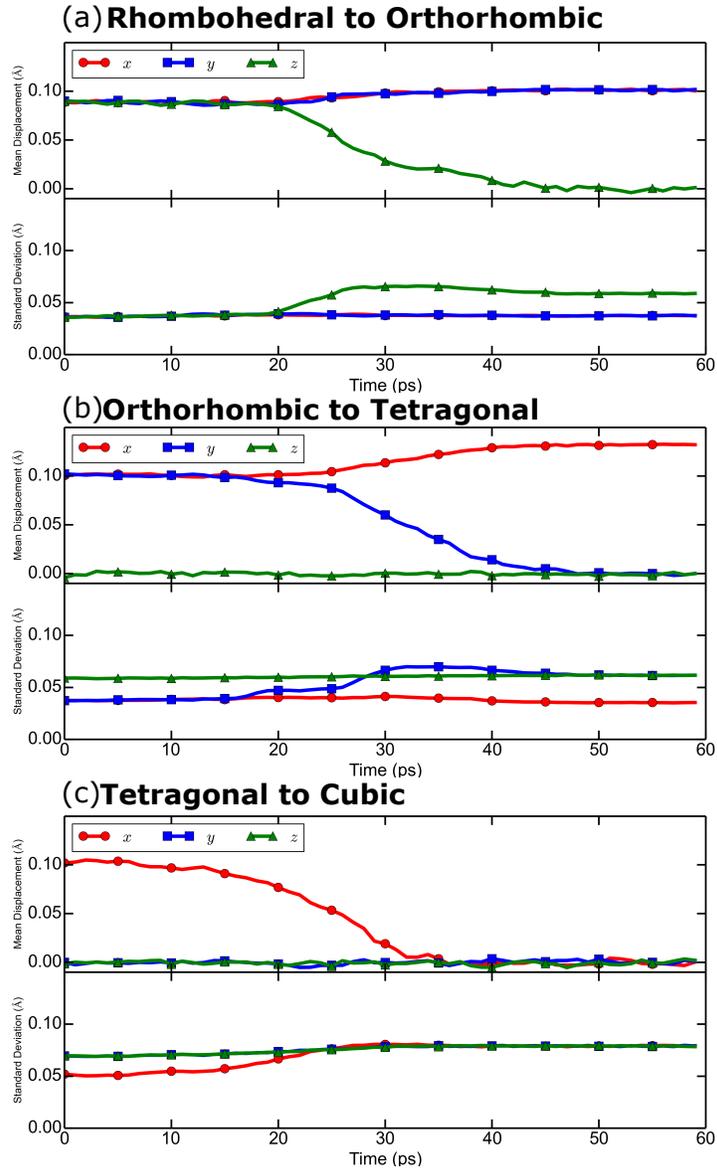}
\caption{The change of the average and standard deviation of the Ti displacement distribution. 
In the standard deviation plot of (b), the green and black lines increase with temperature and are parallel until the transition.}\label{f6}
\end{figure}

To investigate the structural dynamics during phase transitions in more detail, we conducted MD simulations with varying temperatures. 
In three different sets of simulations, the temperatures were increased from 100 K to 110 K (R to O), 110 K to 120 K (O to T) and 155 K to 165 K (T to C) respectively.
The temperature was controlled by the Nos{\'e}--Hoover thermostat with a thermal inertia parameter $M_s$=10 and the 10 K temperature change was accomplished in 60 ps.
We analyze the temperature dependence of Ti displacement distributions along three axes.
The dynamics of Ti displacement distributions during the phase transitions are plotted in FIG.~\ref{f5}.
The time evolution of the average and standard deviation of the Ti displacement distributions are shown in FIG.~\ref{f6}.

Phase transition occurs when one component undergoes polar--nonpolar transition.
The first column (from 100 K to 110 K) shows the changes of Ti displacement distributions during the rhombohedral to orthorhombic phase transition.
In the $x$ and $y$ direction, the averages of the distribution shift up, which is a characteristic of displacive transition. 
Meanwhile, in the $z$ direction, the average becomes zero and the variance becomes significantly larger, indicating that the transition is a mix of displacive and bond--softening characters~\cite{Gaffney07p1444}.
For the orthorhombic to tetragonal phase transition (second column), 
the transition of the $y$ component, which is a polar--nonpolar transition, includes both displacive and bond--softening features.
For the $x$ component, the transition involves both an increase of the average
and a decrease of the standard deviation.
For the $z$ direction, even though the Ti displacement distribution is centered at zero above and below the transition,
the Ti displacements are located closer to zero, indicating an increase in bond hardness. 
From 155 K to 165 K, there is also a bond--hardness--changing transition for the components ($x$ and $y$) with zero averages.
We collectively refer to `bond--softening' and `bond--hardening' as `bond--hardness--changing'.

\begin{figure}[htbp]
\includegraphics[width=12.0cm]{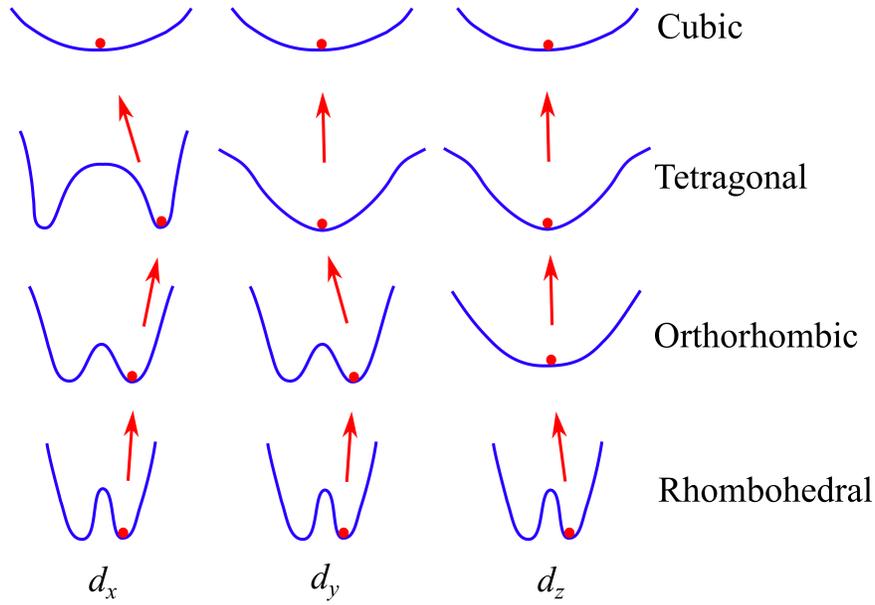}
\caption{Schematic representations of bond--softening, bond--hardening and displacive excitations. 
Two points worth mentioning: (1) For the $x$ component (first column), the minima of the energy profile for the tetragonal phase are further from the center and have a higher curvature, 
compared with those for orthorhombic phase, because the Ti displacement distribution has a larger average and smaller variance;
(2) For the $z$ component (third column), compared with the energy profile for orthorhombic phase, 
the one for the tetragonal phase has a higher curvature at the center (Ti displacement more closely distributed around 0, 
as seen from FIG.~\ref{f5}) and smaller curvature for larger $z$--direction displacements (larger standard deviation, seen from FIG.~\ref{f6} (b)).
}\label{f7}
\end{figure}

Based on the features of the Ti displacement distributions at different phases,
the schematic representation of the thermal excitation between different energy surfaces is presented in FIG.~\ref{f7}.
From our results, the characteristics of BaTiO$_3$ phase transition can be summarized as:
(1) For BaTiO$_3$, the mechanisms of phase transitions include both bond--hardness--changing and displacive transition.
The sudden shifts of the average and standard deviation correspond to displacive with some order--disorder contribution and bond--hardness--changing transitions respectively;
(2) Unlike the conventional understanding that thermal excitation usually causes bond--softening, 
increasing temperature can also cause bond hardening.
The $x$ component of polarization during the orthorhombic to tetragonal transition is an example of this case.
%We collectively refer `bond--softening' and `bond--hardening' as `bond--hardness--changing'; 
(3) When the phase transition occurs, each component of polarization undergoes a change,
even for the component(s) which is(are) non--polar before and after the transition.
The transition(s) that each component undergoes are listed in TABLE~\ref{t6}.

\section{Conclusion}
In this work, we develop a classical atomistic potential for BaTiO$_3$ based on the bond valence model. 
Molecular dynamics simulation with this optimized potential can not only reproduce the temperature--driven phase transitions, 
but can also be a powerful tool in studying the phase transition process with high temporal and spatial resolutions.  
The detailed analysis of the local displacements of Ti atoms reveals that in each phase (including the paraelectric phase),
the majority of Ti atoms are locally displaced, and the phase transitions in BaTiO$_3$ exhibit a mixture of order--disorder and displacive character.
The distribution of Ti displacement is a Gaussian curve or a curve involving a Gaussian and a double peak one. 
Both the average and standard deviation are features of each specific phase, and they can be the order parameters for describing the Gibbs free energy. 
By analyzing the dynamics of Ti displacement distributions during phase transition, we discover several rules of BaTiO$_3$ phase transitions:
the global phase transition is associated with significant changes in each component, even for the components which are nonpolar, 
and the orthorhombic to tetragonal transition exhibits a bond--hardening character in the $z$ component, 
which is opposite to the conventional understanding that temperature increase generally causes bond--softening transition.

\section*{ACKNOWLEDGMENTS}
Y.Q. was supported by the U.S. National Science Foundation, under Grant No. CMMI1334241.
S.L. was supported by the U.S. National Science Foundation, under Grant No. CBET1159736 and Carnegie Institution for Science.
I.G. was supported by the Office of Naval Research, under Grant No. N00014--12--1--1033.
A.M.R. was supported by the Department of Energy, under Grant No. DE--FG02--07ER46431.
Computational support was provided by the High--Performance Computing Modernization Office of the Department of Defense and 
the National Energy Research Scientific Computing Center of the Department of Energy.

\bibliography{rappecites}

\begin{thebibliography}{49}%
\makeatletter
\providecommand \@ifxundefined [1]{%
 \@ifx{#1\undefined}
}%
\providecommand \@ifnum [1]{%
 \ifnum #1\expandafter \@firstoftwo
 \else \expandafter \@secondoftwo
 \fi
}%
\providecommand \@ifx [1]{%
 \ifx #1\expandafter \@firstoftwo
 \else \expandafter \@secondoftwo
 \fi
}%
\providecommand \natexlab [1]{#1}%
\providecommand \enquote  [1]{``#1''}%
\providecommand \bibnamefont  [1]{#1}%
\providecommand \bibfnamefont [1]{#1}%
\providecommand \citenamefont [1]{#1}%
\providecommand \href@noop [0]{\@secondoftwo}%
\providecommand \href [0]{\begingroup \@sanitize@url \@href}%
\providecommand \@href[1]{\@@startlink{#1}\@@href}%
\providecommand \@@href[1]{\endgroup#1\@@endlink}%
\providecommand \@sanitize@url [0]{\catcode `\\12\catcode `\$12\catcode
  `\&12\catcode `\#12\catcode `\^12\catcode `\_12\catcode `\%12\relax}%
\providecommand \@@startlink[1]{}%
\providecommand \@@endlink[0]{}%
\providecommand \url  [0]{\begingroup\@sanitize@url \@url }%
\providecommand \@url [1]{\endgroup\@href {#1}{\urlprefix }}%
\providecommand \urlprefix  [0]{URL }%
\providecommand \Eprint [0]{\href }%
\providecommand \doibase [0]{http://dx.doi.org/}%
\providecommand \selectlanguage [0]{\@gobble}%
\providecommand \bibinfo  [0]{\@secondoftwo}%
\providecommand \bibfield  [0]{\@secondoftwo}%
\providecommand \translation [1]{[#1]}%
\providecommand \BibitemOpen [0]{}%
\providecommand \bibitemStop [0]{}%
\providecommand \bibitemNoStop [0]{.\EOS\space}%
\providecommand \EOS [0]{\spacefactor3000\relax}%
\providecommand \BibitemShut  [1]{\csname bibitem#1\endcsname}%
\let\auto@bib@innerbib\@empty
%</preamble>
\bibitem [{\citenamefont {Yun}\ \emph {et~al.}(2002)\citenamefont {Yun},
  \citenamefont {Urban}, \citenamefont {Gu},\ and\ \citenamefont
  {Park}}]{Yun02p447}%
  \BibitemOpen
  \bibfield  {author} {\bibinfo {author} {\bibfnamefont {W.~S.}\ \bibnamefont
  {Yun}}, \bibinfo {author} {\bibfnamefont {J.~J.}\ \bibnamefont {Urban}},
  \bibinfo {author} {\bibfnamefont {Q.}~\bibnamefont {Gu}}, \ and\ \bibinfo
  {author} {\bibfnamefont {H.}~\bibnamefont {Park}},\ }\href@noop {} {\bibfield
   {journal} {\bibinfo  {journal} {Nano Lett.}\ }\textbf {\bibinfo {volume}
  {2}},\ \bibinfo {pages} {447} (\bibinfo {year} {2002})}\BibitemShut {NoStop}%
\bibitem [{\citenamefont {Chen}\ \emph {et~al.}(2007)\citenamefont {Chen},
  \citenamefont {Chen}, \citenamefont {Chen}, \citenamefont {Yang},\ and\
  \citenamefont {Chang}}]{Chen07p553}%
  \BibitemOpen
  \bibfield  {author} {\bibinfo {author} {\bibfnamefont {K.-H.}\ \bibnamefont
  {Chen}}, \bibinfo {author} {\bibfnamefont {Y.-C.}\ \bibnamefont {Chen}},
  \bibinfo {author} {\bibfnamefont {Z.-S.}\ \bibnamefont {Chen}}, \bibinfo
  {author} {\bibfnamefont {C.-F.}\ \bibnamefont {Yang}}, \ and\ \bibinfo
  {author} {\bibfnamefont {T.-C.}\ \bibnamefont {Chang}},\ }\href@noop {}
  {\bibfield  {journal} {\bibinfo  {journal} {Appl. Phys. A Mater. Sci.}\
  }\textbf {\bibinfo {volume} {89}},\ \bibinfo {pages} {533} (\bibinfo {year}
  {2007})}\BibitemShut {NoStop}%
\bibitem [{\citenamefont {Buessem}\ \emph {et~al.}(1966)\citenamefont
  {Buessem}, \citenamefont {Cross},\ and\ \citenamefont
  {Goswami}}]{Buessem66p33}%
  \BibitemOpen
  \bibfield  {author} {\bibinfo {author} {\bibfnamefont {W.}~\bibnamefont
  {Buessem}}, \bibinfo {author} {\bibfnamefont {L.}~\bibnamefont {Cross}}, \
  and\ \bibinfo {author} {\bibfnamefont {A.}~\bibnamefont {Goswami}},\
  }\href@noop {} {\bibfield  {journal} {\bibinfo  {journal} {J. Am. Ceram.
  Soc.}\ }\textbf {\bibinfo {volume} {49}},\ \bibinfo {pages} {33} (\bibinfo
  {year} {1966})}\BibitemShut {NoStop}%
\bibitem [{\citenamefont {Karaki}\ \emph {et~al.}(2007)\citenamefont {Karaki},
  \citenamefont {Yan},\ and\ \citenamefont {Adachi}}]{Karaki07p7035}%
  \BibitemOpen
  \bibfield  {author} {\bibinfo {author} {\bibfnamefont {T.}~\bibnamefont
  {Karaki}}, \bibinfo {author} {\bibfnamefont {K.}~\bibnamefont {Yan}}, \ and\
  \bibinfo {author} {\bibfnamefont {M.}~\bibnamefont {Adachi}},\ }\href@noop {}
  {\bibfield  {journal} {\bibinfo  {journal} {Jpn. J. of Appl. Phys.}\ }\textbf
  {\bibinfo {volume} {46}},\ \bibinfo {pages} {7035} (\bibinfo {year}
  {2007})}\BibitemShut {NoStop}%
\bibitem [{\citenamefont {Cohen}(1992)}]{Cohen92p65}%
  \BibitemOpen
  \bibfield  {author} {\bibinfo {author} {\bibfnamefont {R.~E.}\ \bibnamefont
  {Cohen}},\ }\href@noop {} {\bibfield  {journal} {\bibinfo  {journal}
  {Ferroelectrics}\ }\textbf {\bibinfo {volume} {136}},\ \bibinfo {pages} {65}
  (\bibinfo {year} {1992})}\BibitemShut {NoStop}%
\bibitem [{\citenamefont {Gonze}\ and\ \citenamefont
  {Lee}(1997)}]{Gonze97p10355}%
  \BibitemOpen
  \bibfield  {author} {\bibinfo {author} {\bibfnamefont {X.}~\bibnamefont
  {Gonze}}\ and\ \bibinfo {author} {\bibfnamefont {C.}~\bibnamefont {Lee}},\
  }\href@noop {} {\bibfield  {journal} {\bibinfo  {journal} {Phys. Rev. B}\
  }\textbf {\bibinfo {volume} {55}},\ \bibinfo {pages} {10355} (\bibinfo {year}
  {1997})}\BibitemShut {NoStop}%
\bibitem [{\citenamefont {Saha}\ \emph {et~al.}(2000)\citenamefont {Saha},
  \citenamefont {Sinha},\ and\ \citenamefont {Mookerjee}}]{Saha00p8828}%
  \BibitemOpen
  \bibfield  {author} {\bibinfo {author} {\bibfnamefont {S.}~\bibnamefont
  {Saha}}, \bibinfo {author} {\bibfnamefont {T.~P.}\ \bibnamefont {Sinha}}, \
  and\ \bibinfo {author} {\bibfnamefont {A.}~\bibnamefont {Mookerjee}},\
  }\href@noop {} {\bibfield  {journal} {\bibinfo  {journal} {Phys. Rev. B}\
  }\textbf {\bibinfo {volume} {62}},\ \bibinfo {pages} {8828} (\bibinfo {year}
  {2000})}\BibitemShut {NoStop}%
\bibitem [{\citenamefont {Kolpak}\ \emph {et~al.}(2008)\citenamefont {Kolpak},
  \citenamefont {Li}, \citenamefont {Shao}, \citenamefont {Rappe},\ and\
  \citenamefont {Bonnell}}]{Kolpak08p036102}%
  \BibitemOpen
  \bibfield  {author} {\bibinfo {author} {\bibfnamefont {A.~M.}\ \bibnamefont
  {Kolpak}}, \bibinfo {author} {\bibfnamefont {D.}~\bibnamefont {Li}}, \bibinfo
  {author} {\bibfnamefont {R.}~\bibnamefont {Shao}}, \bibinfo {author}
  {\bibfnamefont {A.~M.}\ \bibnamefont {Rappe}}, \ and\ \bibinfo {author}
  {\bibfnamefont {D.~A.}\ \bibnamefont {Bonnell}},\ }\href@noop {} {\bibfield
  {journal} {\bibinfo  {journal} {Phys. Rev. Lett.}\ }\textbf {\bibinfo
  {volume} {101}},\ \bibinfo {pages} {036102} (\bibinfo {year}
  {2008})}\BibitemShut {NoStop}%
\bibitem [{\citenamefont {Martirez}\ \emph {et~al.}(2012)\citenamefont
  {Martirez}, \citenamefont {Morales}, \citenamefont {Al-Saidi}, \citenamefont
  {Bonnell},\ and\ \citenamefont {Rappe}}]{Martirez12p256802}%
  \BibitemOpen
  \bibfield  {author} {\bibinfo {author} {\bibfnamefont {J.~M.~P.}\
  \bibnamefont {Martirez}}, \bibinfo {author} {\bibfnamefont {E.~H.}\
  \bibnamefont {Morales}}, \bibinfo {author} {\bibfnamefont {W.~A.}\
  \bibnamefont {Al-Saidi}}, \bibinfo {author} {\bibfnamefont {D.~A.}\
  \bibnamefont {Bonnell}}, \ and\ \bibinfo {author} {\bibfnamefont {A.~M.}\
  \bibnamefont {Rappe}},\ }\href@noop {} {\bibfield  {journal} {\bibinfo
  {journal} {Phys. Rev. Lett.}\ }\textbf {\bibinfo {volume} {109}},\ \bibinfo
  {pages} {256802 1} (\bibinfo {year} {2012})}\BibitemShut {NoStop}%
\bibitem [{\citenamefont {Morales}\ \emph {et~al.}(2014)\citenamefont
  {Morales}, \citenamefont {Martirez}, \citenamefont {Saidi}, \citenamefont
  {Rappe},\ and\ \citenamefont {Bonnell}}]{Morales14p4465}%
  \BibitemOpen
  \bibfield  {author} {\bibinfo {author} {\bibfnamefont {E.~H.}\ \bibnamefont
  {Morales}}, \bibinfo {author} {\bibfnamefont {J.~M.~P.}\ \bibnamefont
  {Martirez}}, \bibinfo {author} {\bibfnamefont {W.~A.}\ \bibnamefont {Saidi}},
  \bibinfo {author} {\bibfnamefont {A.~M.}\ \bibnamefont {Rappe}}, \ and\
  \bibinfo {author} {\bibfnamefont {D.~A.}\ \bibnamefont {Bonnell}},\
  }\href@noop {} {\bibfield  {journal} {\bibinfo  {journal} {ACS nano}\
  }\textbf {\bibinfo {volume} {8}},\ \bibinfo {pages} {4465} (\bibinfo {year}
  {2014})}\BibitemShut {NoStop}%
\bibitem [{\citenamefont {Zhong}\ \emph {et~al.}(1994)\citenamefont {Zhong},
  \citenamefont {Vanderbilt},\ and\ \citenamefont {Rabe}}]{Zhong94p1861}%
  \BibitemOpen
  \bibfield  {author} {\bibinfo {author} {\bibfnamefont {W.}~\bibnamefont
  {Zhong}}, \bibinfo {author} {\bibfnamefont {D.}~\bibnamefont {Vanderbilt}}, \
  and\ \bibinfo {author} {\bibfnamefont {K.~M.}\ \bibnamefont {Rabe}},\
  }\href@noop {} {\bibfield  {journal} {\bibinfo  {journal} {Phys. Rev. Lett.}\
  }\textbf {\bibinfo {volume} {73}},\ \bibinfo {pages} {1861} (\bibinfo {year}
  {1994})}\BibitemShut {NoStop}%
\bibitem [{\citenamefont {Zhong}\ \emph {et~al.}(1995)\citenamefont {Zhong},
  \citenamefont {Vanderbilt},\ and\ \citenamefont {Rabe}}]{Zhong95p6301}%
  \BibitemOpen
  \bibfield  {author} {\bibinfo {author} {\bibfnamefont {W.}~\bibnamefont
  {Zhong}}, \bibinfo {author} {\bibfnamefont {D.}~\bibnamefont {Vanderbilt}}, \
  and\ \bibinfo {author} {\bibfnamefont {K.~M.}\ \bibnamefont {Rabe}},\
  }\href@noop {} {\bibfield  {journal} {\bibinfo  {journal} {Phys. Rev. B}\
  }\textbf {\bibinfo {volume} {52}},\ \bibinfo {pages} {6301} (\bibinfo {year}
  {1995})}\BibitemShut {NoStop}%
\bibitem [{\citenamefont {Nishimatsu}\ \emph {et~al.}(2008)\citenamefont
  {Nishimatsu}, \citenamefont {Waghmare}, \citenamefont {Kawazoe},\ and\
  \citenamefont {Vanderbilt}}]{Nishimatsu08p104104}%
  \BibitemOpen
  \bibfield  {author} {\bibinfo {author} {\bibfnamefont {T.}~\bibnamefont
  {Nishimatsu}}, \bibinfo {author} {\bibfnamefont {U.~V.}\ \bibnamefont
  {Waghmare}}, \bibinfo {author} {\bibfnamefont {Y.}~\bibnamefont {Kawazoe}}, \
  and\ \bibinfo {author} {\bibfnamefont {D.}~\bibnamefont {Vanderbilt}},\
  }\href@noop {} {\bibfield  {journal} {\bibinfo  {journal} {Phys. Rev. B}\
  }\textbf {\bibinfo {volume} {78}},\ \bibinfo {pages} {104104} (\bibinfo
  {year} {2008})}\BibitemShut {NoStop}%
\bibitem [{\citenamefont {Fu}\ and\ \citenamefont
  {Bellaiche}(2003)}]{Fu03p257601}%
  \BibitemOpen
  \bibfield  {author} {\bibinfo {author} {\bibfnamefont {H.}~\bibnamefont
  {Fu}}\ and\ \bibinfo {author} {\bibfnamefont {L.}~\bibnamefont {Bellaiche}},\
  }\href@noop {} {\bibfield  {journal} {\bibinfo  {journal} {Phys. Rev. Lett.}\
  }\textbf {\bibinfo {volume} {91}},\ \bibinfo {pages} {257601} (\bibinfo
  {year} {2003})}\BibitemShut {NoStop}%
\bibitem [{\citenamefont {Tinte}\ \emph {et~al.}(1999)\citenamefont {Tinte},
  \citenamefont {Stachiotti}, \citenamefont {Sepliarsky}, \citenamefont
  {Migoni},\ and\ \citenamefont {Rodriquez}}]{Tinte99p9679}%
  \BibitemOpen
  \bibfield  {author} {\bibinfo {author} {\bibfnamefont {S.}~\bibnamefont
  {Tinte}}, \bibinfo {author} {\bibfnamefont {M.~G.}\ \bibnamefont
  {Stachiotti}}, \bibinfo {author} {\bibfnamefont {M.}~\bibnamefont
  {Sepliarsky}}, \bibinfo {author} {\bibfnamefont {R.~L.}\ \bibnamefont
  {Migoni}}, \ and\ \bibinfo {author} {\bibfnamefont {C.~O.}\ \bibnamefont
  {Rodriquez}},\ }\href@noop {} {\bibfield  {journal} {\bibinfo  {journal} {J.
  Phys.: Condens. Matter}\ }\textbf {\bibinfo {volume} {11}},\ \bibinfo {pages}
  {9679} (\bibinfo {year} {1999})}\BibitemShut {NoStop}%
\bibitem [{\citenamefont {Tinte}\ and\ \citenamefont
  {Stachiotti}(2001)}]{Tinte01p235403}%
  \BibitemOpen
  \bibfield  {author} {\bibinfo {author} {\bibfnamefont {S.}~\bibnamefont
  {Tinte}}\ and\ \bibinfo {author} {\bibfnamefont {M.~G.}\ \bibnamefont
  {Stachiotti}},\ }\href@noop {} {\bibfield  {journal} {\bibinfo  {journal}
  {Phys. Rev. B}\ }\textbf {\bibinfo {volume} {64}},\ \bibinfo {pages} {235403}
  (\bibinfo {year} {2001})}\BibitemShut {NoStop}%
\bibitem [{\citenamefont {Tinte}\ \emph {et~al.}(2000)\citenamefont {Tinte},
  \citenamefont {Stachiotti}, \citenamefont {Sepliarsky}, \citenamefont
  {Migoni},\ and\ \citenamefont {Rodriguez}}]{Tinte00p41}%
  \BibitemOpen
  \bibfield  {author} {\bibinfo {author} {\bibfnamefont {S.}~\bibnamefont
  {Tinte}}, \bibinfo {author} {\bibfnamefont {M.}~\bibnamefont {Stachiotti}},
  \bibinfo {author} {\bibfnamefont {M.}~\bibnamefont {Sepliarsky}}, \bibinfo
  {author} {\bibfnamefont {R.}~\bibnamefont {Migoni}}, \ and\ \bibinfo {author}
  {\bibfnamefont {C.}~\bibnamefont {Rodriguez}},\ }\href@noop {} {\bibfield
  {journal} {\bibinfo  {journal} {Ferroelectrics}\ }\textbf {\bibinfo {volume}
  {237}},\ \bibinfo {pages} {41} (\bibinfo {year} {2000})}\BibitemShut
  {NoStop}%
\bibitem [{\citenamefont {Zhang}\ \emph {et~al.}(2010)\citenamefont {Zhang},
  \citenamefont {Hong}, \citenamefont {Liu},\ and\ \citenamefont
  {Fang}}]{Zhang10p015701}%
  \BibitemOpen
  \bibfield  {author} {\bibinfo {author} {\bibfnamefont {Y.}~\bibnamefont
  {Zhang}}, \bibinfo {author} {\bibfnamefont {J.}~\bibnamefont {Hong}},
  \bibinfo {author} {\bibfnamefont {B.}~\bibnamefont {Liu}}, \ and\ \bibinfo
  {author} {\bibfnamefont {D.}~\bibnamefont {Fang}},\ }\href@noop {} {\bibfield
   {journal} {\bibinfo  {journal} {Nanotechnology}\ }\textbf {\bibinfo {volume}
  {21}},\ \bibinfo {pages} {015701} (\bibinfo {year} {2010})}\BibitemShut
  {NoStop}%
\bibitem [{\citenamefont {Zhang}\ \emph {et~al.}(2009)\citenamefont {Zhang},
  \citenamefont {Hong}, \citenamefont {Liu},\ and\ \citenamefont
  {Fang}}]{Zhang09p405703}%
  \BibitemOpen
  \bibfield  {author} {\bibinfo {author} {\bibfnamefont {Y.}~\bibnamefont
  {Zhang}}, \bibinfo {author} {\bibfnamefont {J.}~\bibnamefont {Hong}},
  \bibinfo {author} {\bibfnamefont {B.}~\bibnamefont {Liu}}, \ and\ \bibinfo
  {author} {\bibfnamefont {D.}~\bibnamefont {Fang}},\ }\href@noop {} {\bibfield
   {journal} {\bibinfo  {journal} {Nanotechnology}\ }\textbf {\bibinfo {volume}
  {20}},\ \bibinfo {pages} {405703} (\bibinfo {year} {2009})}\BibitemShut
  {NoStop}%
\bibitem [{\citenamefont {Grinberg}\ \emph {et~al.}(2002)\citenamefont
  {Grinberg}, \citenamefont {Cooper},\ and\ \citenamefont
  {Rappe}}]{Grinberg02p909}%
  \BibitemOpen
  \bibfield  {author} {\bibinfo {author} {\bibfnamefont {I.}~\bibnamefont
  {Grinberg}}, \bibinfo {author} {\bibfnamefont {V.~R.}\ \bibnamefont
  {Cooper}}, \ and\ \bibinfo {author} {\bibfnamefont {A.~M.}\ \bibnamefont
  {Rappe}},\ }\href@noop {} {\bibfield  {journal} {\bibinfo  {journal}
  {Nature}\ }\textbf {\bibinfo {volume} {419}},\ \bibinfo {pages} {909}
  (\bibinfo {year} {2002})}\BibitemShut {NoStop}%
\bibitem [{\citenamefont {Shin}\ \emph {et~al.}(2008)\citenamefont {Shin},
  \citenamefont {Son}, \citenamefont {Lee}, \citenamefont {Grinberg},\ and\
  \citenamefont {Rappe}}]{Shin08p015224}%
  \BibitemOpen
  \bibfield  {author} {\bibinfo {author} {\bibfnamefont {Y.-H.}\ \bibnamefont
  {Shin}}, \bibinfo {author} {\bibfnamefont {J.-Y.}\ \bibnamefont {Son}},
  \bibinfo {author} {\bibfnamefont {B.-J.}\ \bibnamefont {Lee}}, \bibinfo
  {author} {\bibfnamefont {I.}~\bibnamefont {Grinberg}}, \ and\ \bibinfo
  {author} {\bibfnamefont {A.~M.}\ \bibnamefont {Rappe}},\ }\href@noop {}
  {\bibfield  {journal} {\bibinfo  {journal} {J. Phys.: Condens. Matter}\
  }\textbf {\bibinfo {volume} {20}},\ \bibinfo {pages} {015224} (\bibinfo
  {year} {2008})}\BibitemShut {NoStop}%
\bibitem [{\citenamefont {Shin}\ \emph {et~al.}(2005)\citenamefont {Shin},
  \citenamefont {Cooper}, \citenamefont {Grinberg},\ and\ \citenamefont
  {Rappe}}]{Shin05p054104}%
  \BibitemOpen
  \bibfield  {author} {\bibinfo {author} {\bibfnamefont {Y.-H.}\ \bibnamefont
  {Shin}}, \bibinfo {author} {\bibfnamefont {V.~R.}\ \bibnamefont {Cooper}},
  \bibinfo {author} {\bibfnamefont {I.}~\bibnamefont {Grinberg}}, \ and\
  \bibinfo {author} {\bibfnamefont {A.~M.}\ \bibnamefont {Rappe}},\ }\href
  {http://dx.doi.org/10.1103/PhysRevB.71.054104} {\bibfield  {journal}
  {\bibinfo  {journal} {Phys. Rev. B}\ }\textbf {\bibinfo {volume} {71}},\
  \bibinfo {pages} {054104} (\bibinfo {year} {2005})}\BibitemShut {NoStop}%
\bibitem [{\citenamefont {Liu}\ \emph {et~al.}(2013{\natexlab{a}})\citenamefont
  {Liu}, \citenamefont {Grinberg},\ and\ \citenamefont {Rappe}}]{Liu13p102202}%
  \BibitemOpen
  \bibfield  {author} {\bibinfo {author} {\bibfnamefont {S.}~\bibnamefont
  {Liu}}, \bibinfo {author} {\bibfnamefont {I.}~\bibnamefont {Grinberg}}, \
  and\ \bibinfo {author} {\bibfnamefont {A.~M.}\ \bibnamefont {Rappe}},\
  }\href@noop {} {\bibfield  {journal} {\bibinfo  {journal} {J. Physics.:
  Condens. Matter}\ }\textbf {\bibinfo {volume} {25}},\ \bibinfo {pages}
  {102202} (\bibinfo {year} {2013}{\natexlab{a}})}\BibitemShut {NoStop}%
\bibitem [{\citenamefont {Liu}\ \emph {et~al.}(2013{\natexlab{b}})\citenamefont
  {Liu}, \citenamefont {Grinberg}, \citenamefont {Takenaka},\ and\
  \citenamefont {Rappe}}]{Liu13p104102}%
  \BibitemOpen
  \bibfield  {author} {\bibinfo {author} {\bibfnamefont {S.}~\bibnamefont
  {Liu}}, \bibinfo {author} {\bibfnamefont {I.}~\bibnamefont {Grinberg}},
  \bibinfo {author} {\bibfnamefont {H.}~\bibnamefont {Takenaka}}, \ and\
  \bibinfo {author} {\bibfnamefont {A.~M.}\ \bibnamefont {Rappe}},\ }\href@noop
  {} {\bibfield  {journal} {\bibinfo  {journal} {Phys. Rev. B}\ }\textbf
  {\bibinfo {volume} {88}},\ \bibinfo {pages} {104102} (\bibinfo {year}
  {2013}{\natexlab{b}})}\BibitemShut {NoStop}%
\bibitem [{\citenamefont {Chen}\ \emph {et~al.}(2015)\citenamefont {Chen},
  \citenamefont {Goodfellow}, \citenamefont {Liu}, \citenamefont {Grinberg},
  \citenamefont {Hoffmann}, \citenamefont {Damodaran}, \citenamefont {Zhu},
  \citenamefont {Zalden}, \citenamefont {Zhang}, \citenamefont {Takeuchi},
  \citenamefont {Rappe}, \citenamefont {Martin}, \citenamefont {Wen},\ and\
  \citenamefont {Lindenberg}}]{Chen15p6371}%
  \BibitemOpen
  \bibfield  {author} {\bibinfo {author} {\bibfnamefont {F.}~\bibnamefont
  {Chen}}, \bibinfo {author} {\bibfnamefont {J.}~\bibnamefont {Goodfellow}},
  \bibinfo {author} {\bibfnamefont {S.}~\bibnamefont {Liu}}, \bibinfo {author}
  {\bibfnamefont {I.}~\bibnamefont {Grinberg}}, \bibinfo {author}
  {\bibfnamefont {M.~C.}\ \bibnamefont {Hoffmann}}, \bibinfo {author}
  {\bibfnamefont {A.~R.}\ \bibnamefont {Damodaran}}, \bibinfo {author}
  {\bibfnamefont {Y.}~\bibnamefont {Zhu}}, \bibinfo {author} {\bibfnamefont
  {P.}~\bibnamefont {Zalden}}, \bibinfo {author} {\bibfnamefont
  {X.}~\bibnamefont {Zhang}}, \bibinfo {author} {\bibfnamefont
  {I.}~\bibnamefont {Takeuchi}}, \bibinfo {author} {\bibfnamefont {A.~M.}\
  \bibnamefont {Rappe}}, \bibinfo {author} {\bibfnamefont {L.~W.}\ \bibnamefont
  {Martin}}, \bibinfo {author} {\bibfnamefont {H.}~\bibnamefont {Wen}}, \ and\
  \bibinfo {author} {\bibfnamefont {A.~M.}\ \bibnamefont {Lindenberg}},\
  }\href@noop {} {\bibfield  {journal} {\bibinfo  {journal} {Adv. Mater.}\
  }\textbf {\bibinfo {volume} {27}},\ \bibinfo {pages} {6371} (\bibinfo {year}
  {2015})}\BibitemShut {NoStop}%
\bibitem [{\citenamefont {Liu}\ \emph {et~al.}(2013{\natexlab{c}})\citenamefont
  {Liu}, \citenamefont {Grinberg},\ and\ \citenamefont {Rappe}}]{Liu13p232907}%
  \BibitemOpen
  \bibfield  {author} {\bibinfo {author} {\bibfnamefont {S.}~\bibnamefont
  {Liu}}, \bibinfo {author} {\bibfnamefont {I.}~\bibnamefont {Grinberg}}, \
  and\ \bibinfo {author} {\bibfnamefont {A.~M.}\ \bibnamefont {Rappe}},\
  }\href@noop {} {\bibfield  {journal} {\bibinfo  {journal} {Appl. Phys.
  Lett.}\ }\textbf {\bibinfo {volume} {103}},\ \bibinfo {pages} {232907}
  (\bibinfo {year} {2013}{\natexlab{c}})}\BibitemShut {NoStop}%
\bibitem [{\citenamefont {Xu}\ \emph {et~al.}(2015)\citenamefont {Xu},
  \citenamefont {Liu}, \citenamefont {Grinberg}, \citenamefont {Karthik},
  \citenamefont {Damodaran}, \citenamefont {Rappe},\ and\ \citenamefont
  {Martin}}]{Xu15p79}%
  \BibitemOpen
  \bibfield  {author} {\bibinfo {author} {\bibfnamefont {R.}~\bibnamefont
  {Xu}}, \bibinfo {author} {\bibfnamefont {S.}~\bibnamefont {Liu}}, \bibinfo
  {author} {\bibfnamefont {I.}~\bibnamefont {Grinberg}}, \bibinfo {author}
  {\bibfnamefont {J.}~\bibnamefont {Karthik}}, \bibinfo {author} {\bibfnamefont
  {A.~R.}\ \bibnamefont {Damodaran}}, \bibinfo {author} {\bibfnamefont {A.~M.}\
  \bibnamefont {Rappe}}, \ and\ \bibinfo {author} {\bibfnamefont {L.~W.}\
  \bibnamefont {Martin}},\ }\href@noop {} {\bibfield  {journal} {\bibinfo
  {journal} {Nat. Mater.}\ }\textbf {\bibinfo {volume} {14}},\ \bibinfo {pages}
  {79} (\bibinfo {year} {2015})}\BibitemShut {NoStop}%
\bibitem [{\citenamefont {Kwei}\ \emph {et~al.}(1993)\citenamefont {Kwei},
  \citenamefont {Lawson}, \citenamefont {Billinge},\ and\ \citenamefont
  {Cheong}}]{Kwei93p2368}%
  \BibitemOpen
  \bibfield  {author} {\bibinfo {author} {\bibfnamefont {G.}~\bibnamefont
  {Kwei}}, \bibinfo {author} {\bibfnamefont {A.}~\bibnamefont {Lawson}},
  \bibinfo {author} {\bibfnamefont {S.}~\bibnamefont {Billinge}}, \ and\
  \bibinfo {author} {\bibfnamefont {S.}~\bibnamefont {Cheong}},\ }\href@noop {}
  {\bibfield  {journal} {\bibinfo  {journal} {J. Phys. Chem.}\ }\textbf
  {\bibinfo {volume} {97}},\ \bibinfo {pages} {2368} (\bibinfo {year}
  {1993})}\BibitemShut {NoStop}%
\bibitem [{\citenamefont {Di\'{e}guez}\ \emph {et~al.}(2004)\citenamefont
  {Di\'{e}guez}, \citenamefont {Tinte}, \citenamefont {Antons}, \citenamefont
  {Bungaro}, \citenamefont {Neaton}, \citenamefont {Rabe},\ and\ \citenamefont
  {Vanderbilt}}]{Dieguez04p212101}%
  \BibitemOpen
  \bibfield  {author} {\bibinfo {author} {\bibfnamefont {O.}~\bibnamefont
  {Di\'{e}guez}}, \bibinfo {author} {\bibfnamefont {S.}~\bibnamefont {Tinte}},
  \bibinfo {author} {\bibfnamefont {A.}~\bibnamefont {Antons}}, \bibinfo
  {author} {\bibfnamefont {C.}~\bibnamefont {Bungaro}}, \bibinfo {author}
  {\bibfnamefont {J.~B.}\ \bibnamefont {Neaton}}, \bibinfo {author}
  {\bibfnamefont {K.~M.}\ \bibnamefont {Rabe}}, \ and\ \bibinfo {author}
  {\bibfnamefont {D.}~\bibnamefont {Vanderbilt}},\ }\href@noop {} {\bibfield
  {journal} {\bibinfo  {journal} {Phys. Rev. B}\ }\textbf {\bibinfo {volume}
  {69}},\ \bibinfo {pages} {212101 1} (\bibinfo {year} {2004})}\BibitemShut
  {NoStop}%
\bibitem [{\citenamefont {Von~Hippel}(1950)}]{Von50p221}%
  \BibitemOpen
  \bibfield  {author} {\bibinfo {author} {\bibfnamefont {A.}~\bibnamefont
  {Von~Hippel}},\ }\href@noop {} {\bibfield  {journal} {\bibinfo  {journal}
  {Rev. Mod. Phys.}\ }\textbf {\bibinfo {volume} {22}},\ \bibinfo {pages} {221}
  (\bibinfo {year} {1950})}\BibitemShut {NoStop}%
\bibitem [{\citenamefont {Brown}(2009)}]{Brown09p6858}%
  \BibitemOpen
  \bibfield  {author} {\bibinfo {author} {\bibfnamefont {I.~D.}\ \bibnamefont
  {Brown}},\ }\href@noop {} {\bibfield  {journal} {\bibinfo  {journal} {Chem.
  Rev.}\ }\textbf {\bibinfo {volume} {109}},\ \bibinfo {pages} {6858} (\bibinfo
  {year} {2009})}\BibitemShut {NoStop}%
\bibitem [{\citenamefont {Brown}\ and\ \citenamefont
  {Shannon}(1973)}]{Brown73p266}%
  \BibitemOpen
  \bibfield  {author} {\bibinfo {author} {\bibfnamefont {I.}~\bibnamefont
  {Brown}}\ and\ \bibinfo {author} {\bibfnamefont {R.}~\bibnamefont
  {Shannon}},\ }\href@noop {} {\bibfield  {journal} {\bibinfo  {journal} {Acta
  Crystallogr. A}\ }\textbf {\bibinfo {volume} {29}},\ \bibinfo {pages} {266}
  (\bibinfo {year} {1973})}\BibitemShut {NoStop}%
\bibitem [{\citenamefont {Brown}\ and\ \citenamefont
  {Wu}(1976)}]{Brown76p1957}%
  \BibitemOpen
  \bibfield  {author} {\bibinfo {author} {\bibfnamefont {I.}~\bibnamefont
  {Brown}}\ and\ \bibinfo {author} {\bibfnamefont {K.~K.}\ \bibnamefont {Wu}},\
  }\href@noop {} {\bibfield  {journal} {\bibinfo  {journal} {Acta Crystallogr.
  B}\ }\textbf {\bibinfo {volume} {32}},\ \bibinfo {pages} {1957} (\bibinfo
  {year} {1976})}\BibitemShut {NoStop}%
\bibitem [{\citenamefont {Finnis}\ and\ \citenamefont
  {Sinclair}(1984)}]{Finnis84p45}%
  \BibitemOpen
  \bibfield  {author} {\bibinfo {author} {\bibfnamefont {M.}~\bibnamefont
  {Finnis}}\ and\ \bibinfo {author} {\bibfnamefont {J.}~\bibnamefont
  {Sinclair}},\ }\href@noop {} {\bibfield  {journal} {\bibinfo  {journal}
  {Philos. Mag. A}\ }\textbf {\bibinfo {volume} {50}},\ \bibinfo {pages} {45}
  (\bibinfo {year} {1984})}\BibitemShut {NoStop}%
\bibitem [{\citenamefont {Harvey}\ \emph {et~al.}(2006)\citenamefont {Harvey},
  \citenamefont {Baggio},\ and\ \citenamefont {Baggio}}]{Harvey06p1038}%
  \BibitemOpen
  \bibfield  {author} {\bibinfo {author} {\bibfnamefont {M.~A.}\ \bibnamefont
  {Harvey}}, \bibinfo {author} {\bibfnamefont {S.}~\bibnamefont {Baggio}}, \
  and\ \bibinfo {author} {\bibfnamefont {R.}~\bibnamefont {Baggio}},\
  }\href@noop {} {\bibfield  {journal} {\bibinfo  {journal} {Acta Crystallogr.
  B}\ }\textbf {\bibinfo {volume} {62}},\ \bibinfo {pages} {1038} (\bibinfo
  {year} {2006})}\BibitemShut {NoStop}%
\bibitem [{\citenamefont {Giannozzi}\ \emph {et~al.}(2009)\citenamefont
  {Giannozzi}, \citenamefont {Baroni}, \citenamefont {Bonini}, \citenamefont
  {Calandra} \emph {et~al.}}]{Giannozzi09p395502etal}%
  \BibitemOpen
  \bibfield  {author} {\bibinfo {author} {\bibfnamefont {P.}~\bibnamefont
  {Giannozzi}}, \bibinfo {author} {\bibfnamefont {S.}~\bibnamefont {Baroni}},
  \bibinfo {author} {\bibfnamefont {N.}~\bibnamefont {Bonini}}, \bibinfo
  {author} {\bibfnamefont {M.}~\bibnamefont {Calandra}},  \emph {et~al.},\
  }\href@noop {} {\bibfield  {journal} {\bibinfo  {journal} {J. Phys.: Condens.
  Matter}\ }\textbf {\bibinfo {volume} {21}},\ \bibinfo {pages} {395502}
  (\bibinfo {year} {2009})}\BibitemShut {NoStop}%
\bibitem [{\citenamefont {Perdew}\ \emph {et~al.}(2008)\citenamefont {Perdew},
  \citenamefont {Ruzsinszky}, \citenamefont {Csonka}, \citenamefont {Vydrov},
  \citenamefont {Scuseria}, \citenamefont {Constantin}, \citenamefont {Zhou},\
  and\ \citenamefont {Burke}}]{Perdew08p136406}%
  \BibitemOpen
  \bibfield  {author} {\bibinfo {author} {\bibfnamefont {J.~P.}\ \bibnamefont
  {Perdew}}, \bibinfo {author} {\bibfnamefont {A.}~\bibnamefont {Ruzsinszky}},
  \bibinfo {author} {\bibfnamefont {G.~I.}\ \bibnamefont {Csonka}}, \bibinfo
  {author} {\bibfnamefont {O.~A.}\ \bibnamefont {Vydrov}}, \bibinfo {author}
  {\bibfnamefont {G.~E.}\ \bibnamefont {Scuseria}}, \bibinfo {author}
  {\bibfnamefont {L.~A.}\ \bibnamefont {Constantin}}, \bibinfo {author}
  {\bibfnamefont {X.}~\bibnamefont {Zhou}}, \ and\ \bibinfo {author}
  {\bibfnamefont {K.}~\bibnamefont {Burke}},\ }\href@noop {} {\bibfield
  {journal} {\bibinfo  {journal} {Phys. Rev. Lett.}\ }\textbf {\bibinfo
  {volume} {100}},\ \bibinfo {pages} {136406} (\bibinfo {year}
  {2008})}\BibitemShut {NoStop}%
\bibitem [{Opi()}]{Opium}%
  \BibitemOpen
  \href@noop {} {}\bibinfo {howpublished}
  {http://opium.sourceforge.net}\BibitemShut {NoStop}%
\bibitem [{\citenamefont {Monkhorst}\ and\ \citenamefont
  {Pack}(1976)}]{Monkhorst76p5188}%
  \BibitemOpen
  \bibfield  {author} {\bibinfo {author} {\bibfnamefont {H.~J.}\ \bibnamefont
  {Monkhorst}}\ and\ \bibinfo {author} {\bibfnamefont {J.~D.}\ \bibnamefont
  {Pack}},\ }\href@noop {} {\bibfield  {journal} {\bibinfo  {journal} {Phys.
  Rev. B}\ }\textbf {\bibinfo {volume} {13}},\ \bibinfo {pages} {5188}
  (\bibinfo {year} {1976})}\BibitemShut {NoStop}%
\bibitem [{\citenamefont {Parrinello}\ and\ \citenamefont
  {Rahman}(1980)}]{Parrinello80p1196}%
  \BibitemOpen
  \bibfield  {author} {\bibinfo {author} {\bibfnamefont {M.}~\bibnamefont
  {Parrinello}}\ and\ \bibinfo {author} {\bibfnamefont {A.}~\bibnamefont
  {Rahman}},\ }\href@noop {} {\bibfield  {journal} {\bibinfo  {journal} {Phys.
  Rev. Lett.}\ }\textbf {\bibinfo {volume} {45}},\ \bibinfo {pages} {1196}
  (\bibinfo {year} {1980})}\BibitemShut {NoStop}%
\bibitem [{\citenamefont {Sepliarsky}\ \emph {et~al.}(2005)\citenamefont
  {Sepliarsky}, \citenamefont {Stachiotti},\ and\ \citenamefont
  {Migoni}}]{Sepliarsky05p014110}%
  \BibitemOpen
  \bibfield  {author} {\bibinfo {author} {\bibfnamefont {M.}~\bibnamefont
  {Sepliarsky}}, \bibinfo {author} {\bibfnamefont {M.~G.}\ \bibnamefont
  {Stachiotti}}, \ and\ \bibinfo {author} {\bibfnamefont {R.~L.}\ \bibnamefont
  {Migoni}},\ }\href@noop {} {\bibfield  {journal} {\bibinfo  {journal} {Phys.
  Rev. B}\ }\textbf {\bibinfo {volume} {72}},\ \bibinfo {pages} {014110}
  (\bibinfo {year} {2005})}\BibitemShut {NoStop}%
\bibitem [{\citenamefont {Cohen}\ and\ \citenamefont
  {Krakauer}(1990)}]{Cohen90p6416}%
  \BibitemOpen
  \bibfield  {author} {\bibinfo {author} {\bibfnamefont {R.~E.}\ \bibnamefont
  {Cohen}}\ and\ \bibinfo {author} {\bibfnamefont {H.}~\bibnamefont
  {Krakauer}},\ }\href@noop {} {\bibfield  {journal} {\bibinfo  {journal}
  {Phys. Rev. B}\ }\textbf {\bibinfo {volume} {42}},\ \bibinfo {pages} {6416}
  (\bibinfo {year} {1990})}\BibitemShut {NoStop}%
\bibitem [{\citenamefont {Gaudon}(2015)}]{Gaudon15p6}%
  \BibitemOpen
  \bibfield  {author} {\bibinfo {author} {\bibfnamefont {M.}~\bibnamefont
  {Gaudon}},\ }\href@noop {} {\bibfield  {journal} {\bibinfo  {journal}
  {Polyhedron}\ }\textbf {\bibinfo {volume} {88}},\ \bibinfo {pages} {6}
  (\bibinfo {year} {2015})}\BibitemShut {NoStop}%
\bibitem [{\citenamefont {Ehses}\ \emph {et~al.}(1981)\citenamefont {Ehses},
  \citenamefont {Bock},\ and\ \citenamefont {Fischer}}]{Ehses81p507}%
  \BibitemOpen
  \bibfield  {author} {\bibinfo {author} {\bibfnamefont {K.~H.}\ \bibnamefont
  {Ehses}}, \bibinfo {author} {\bibfnamefont {H.}~\bibnamefont {Bock}}, \ and\
  \bibinfo {author} {\bibfnamefont {K.}~\bibnamefont {Fischer}},\ }\href@noop
  {} {\bibfield  {journal} {\bibinfo  {journal} {Ferroelectrics}\ }\textbf
  {\bibinfo {volume} {37}},\ \bibinfo {pages} {507} (\bibinfo {year}
  {1981})}\BibitemShut {NoStop}%
\bibitem [{\citenamefont {Jun}\ \emph {et~al.}(1988)\citenamefont {Jun},
  \citenamefont {Chan-Gao}, \citenamefont {Qi},\ and\ \citenamefont
  {Duan}}]{Jun88p2255}%
  \BibitemOpen
  \bibfield  {author} {\bibinfo {author} {\bibfnamefont {C.}~\bibnamefont
  {Jun}}, \bibinfo {author} {\bibfnamefont {F.}~\bibnamefont {Chan-Gao}},
  \bibinfo {author} {\bibfnamefont {L.}~\bibnamefont {Qi}}, \ and\ \bibinfo
  {author} {\bibfnamefont {F.}~\bibnamefont {Duan}},\ }\href@noop {} {\bibfield
   {journal} {\bibinfo  {journal} {J. Phys. C: Solid State Phys.}\ }\textbf
  {\bibinfo {volume} {21}},\ \bibinfo {pages} {2255} (\bibinfo {year}
  {1988})}\BibitemShut {NoStop}%
\bibitem [{\citenamefont {Comes}\ \emph {et~al.}(1968)\citenamefont {Comes},
  \citenamefont {Lambert},\ and\ \citenamefont {Guinier}}]{Comes68p715}%
  \BibitemOpen
  \bibfield  {author} {\bibinfo {author} {\bibfnamefont {R.}~\bibnamefont
  {Comes}}, \bibinfo {author} {\bibfnamefont {M.}~\bibnamefont {Lambert}}, \
  and\ \bibinfo {author} {\bibfnamefont {A.}~\bibnamefont {Guinier}},\
  }\href@noop {} {\bibfield  {journal} {\bibinfo  {journal} {Solid State
  Commun.}\ }\textbf {\bibinfo {volume} {6}},\ \bibinfo {pages} {715} (\bibinfo
  {year} {1968})}\BibitemShut {NoStop}%
\bibitem [{\citenamefont {Itoh}\ \emph {et~al.}(1985)\citenamefont {Itoh},
  \citenamefont {Zeng}, \citenamefont {Nakamura},\ and\ \citenamefont
  {Mishima}}]{Itoh85p29}%
  \BibitemOpen
  \bibfield  {author} {\bibinfo {author} {\bibfnamefont {K.}~\bibnamefont
  {Itoh}}, \bibinfo {author} {\bibfnamefont {L.}~\bibnamefont {Zeng}}, \bibinfo
  {author} {\bibfnamefont {E.}~\bibnamefont {Nakamura}}, \ and\ \bibinfo
  {author} {\bibfnamefont {N.}~\bibnamefont {Mishima}},\ }\href@noop {}
  {\bibfield  {journal} {\bibinfo  {journal} {Ferroelectrics}\ }\textbf
  {\bibinfo {volume} {63}},\ \bibinfo {pages} {29} (\bibinfo {year}
  {1985})}\BibitemShut {NoStop}%
\bibitem [{\citenamefont {Stern}(2004)}]{Stern04p037601}%
  \BibitemOpen
  \bibfield  {author} {\bibinfo {author} {\bibfnamefont {E.~A.}\ \bibnamefont
  {Stern}},\ }\href@noop {} {\bibfield  {journal} {\bibinfo  {journal} {Phys.
  Rev. Lett.}\ }\textbf {\bibinfo {volume} {93}},\ \bibinfo {pages} {037601}
  (\bibinfo {year} {2004})}\BibitemShut {NoStop}%
\bibitem [{\citenamefont {Gaffney}\ and\ \citenamefont
  {Chapman}(2007)}]{Gaffney07p1444}%
  \BibitemOpen
  \bibfield  {author} {\bibinfo {author} {\bibfnamefont {K.~J.}\ \bibnamefont
  {Gaffney}}\ and\ \bibinfo {author} {\bibfnamefont {H.~N.}\ \bibnamefont
  {Chapman}},\ }\href@noop {} {\bibfield  {journal} {\bibinfo  {journal}
  {Science}\ }\textbf {\bibinfo {volume} {316}},\ \bibinfo {pages} {1444}
  (\bibinfo {year} {2007})}\BibitemShut {NoStop}%
\end{thebibliography}%

\newpage
\begin{table}
 \begin{tabular}{ P{1cm} P{1.2cm} P{1.2cm} P{1.5cm} P{1.5cm} P{1.5cm} P{1.5cm} P{1.5cm} P{1.5cm} P{1.2cm} P{1.5cm} }
  \hline
  \hline
  &    &    &    &    &    &    &  $B_{\beta\beta^{\prime}}\ \left({\rm{\AA}}\right)$  &    &    & \\
  \cline{7-9}
  & $r_{0,{\beta}\rm{O}}$ &  $C_{0,\beta\rm{O}}$  &  $q_\beta$(e)  & $S_\beta$(eV)  & $D_\beta$   &   Ba  &  Ti  &  O  &  $V_{0,\beta}$  & ${\bf{W}}_{0,\beta}$ \\   
  \hline
Ba & 2.290 & 8.94 & 1.34730 & 0.59739 & 0.08429 & 2.44805 & 2.32592 & 1.98792 & 2.0 & 0.11561 \\
Ti & 1.798 & 5.20 & 1.28905 & 0.16533 & 0.82484 &         & 2.73825 & 1.37741 & 4.0 & 0.39437 \\
O  &       &      & -0.87878 & 0.93063 & 0.28006 &        &         & 1.99269 & 2.0 & 0.31651 \\
\hline
\hline
 \end{tabular}
 \caption{Optimized force field for BaTiO$_3$. The angle constant $k=6.1$ meV/(deg)$^2$.}\label{t1}
\end{table}

\begin{table}
 \begin{tabular}{ P{2.5cm}P{1.5cm}P{1.5cm}P{1.5cm} }
  \hline\hline
             &  R--O  &  O--T  & T--C   \\  
   \hline       
 BV model &  100 K &  110 K & 160 K  \\
  \hline
shell model  &  80 K  &  120 K & 170 K  \\
  \hline
  \hline
 \end{tabular}
 \caption{Comparison of the phase transition temperatures given by the BV model and the shell model~\cite{Tinte99p9679}.}\label{t2}
\end{table}

\begin{table}
 \begin{tabular}{ P{3.5cm} P{3.5cm} P{3.5cm} P{3.5cm} }
  \hline
  \hline
  Lattice constant & MD (\AA) & DFT (\AA) & error \\
  \hline
   Rhombohedral    &     &     &    \\     
   $a=b=c$  &  4.036  &  4.024 &  0.30$\%$\\ 
     \hline
   Orthorhombic    &     &     &    \\     
   $a$  &  3.997  &  3.977 &  0.50$\%$\\
   $b=c$  &  4.059  &  4.046 &  0.32$\%$\\
  \hline
      Tetragonal    &     &     &    \\     
   $a=b$  &  4.005  &  3.985 &  0.50$\%$\\
   $c$  &  4.109  &  4.089 &  0.49$\%$\\
   \hline
       Cubic    &     &     &    \\     
   $a=b=c$  &  4.037  &  4.002 &  0.87$\%$\\
     \hline
     \hline
 \end{tabular}
 \caption{Comparison of lattice constants of BaTiO$_3$ given by MD simulations with BV model potential and PBEsol DFT calculations. 
 For MD simulation, lattice constants of rhombohedral, orthorhombic, tetragonal and cubic phases are obtained at 5 K, 105 K, 120 K and 165 K respectively.
 Since DFT neglects thermal expansion, the results given by MD simulations, which are larger but less than 1\%, 
 demonstrate that this set of potential can predicts the lattice constants of BaTiO$_3$ quite well. 
 }\label{t3}
\end{table}

%\begin{table}
% \begin{tabular}{ |P{3cm}P{3.cm}P{3.cm}P{3.cm}P{3.cm}P{3.cm}| }
% \hline
%  \multicolumn{1}{|c|}{} & \multicolumn{1}{c|}{Phase of}& \multicolumn{4}{c|}{Micro--composition} \\
% \cline{3-6}
%  \multicolumn{1}{|c|}{Temperature} & \multicolumn{1}{c|}{the supercell}    &  \multicolumn{1}{c|}{Rhombohedral} &   \multicolumn{1}{c|}{Orthorhombic} &  \multicolumn{1}{c|}{$\ $Tetragonal$\ $} & \multicolumn{1}{c|}{$\ \ \ \ $Cubic$\ \ \ \ $}  \\
%   \hline          
% \multicolumn{1}{|c|}{30 K} & \multicolumn{1}{c|}{Rhombohedral}  &   \multicolumn{1}{c|}{99.279\%} &   \multicolumn{1}{c|}{0.721\%} &  \multicolumn{1}{c|}{0.000\%} & \multicolumn{1}{c|}{0.000\%}  \\
%  \hline          
% \multicolumn{1}{|c|}{70 K} & \multicolumn{1}{c|}{Rhombohedral}  &   \multicolumn{1}{c|}{68.321\%} &   \multicolumn{1}{c|}{30.861\%} &  \multicolumn{1}{c|}{0.767\%} & \multicolumn{1}{c|}{0.051\%}  \\
%    \hline   
% \multicolumn{1}{|c|}{110 K} & \multicolumn{1}{c|}{Orthorhombic}  &   \multicolumn{1}{c|}{18.943\%} &   \multicolumn{1}{c|}{67.575\%} &  \multicolumn{1}{c|}{11.010\%} & \multicolumn{1}{c|}{2.472\%}  \\
% \hline 
% \multicolumn{1}{|c|}{150 K} & \multicolumn{1}{c|}{Tetragonal}  &   \multicolumn{1}{c|}{10.665\%} &   \multicolumn{1}{c|}{51.439\%} &  \multicolumn{1}{c|}{28.492\%} & \multicolumn{1}{c|}{9.404\%}  \\
%\hline 
% \multicolumn{1}{|c|}{190 K} & \multicolumn{1}{c|}{Cubic}  &   \multicolumn{1}{c|}{9.646\%} &   \multicolumn{1}{c|}{44.624\%} &  \multicolumn{1}{c|}{21.843\%} & \multicolumn{1}{c|}{23.887\%}  \\
%  \hline
%\end{tabular}
% \caption{Compositions of each phase.}\label{t4}
%\end{table}

\begin{table}
 \begin{tabular}{ P{3cm}P{3cm}P{3cm}P{3cm}P{3cm} }
  \hline
  \hline
             &  Rhombohedral  &  Orthorhombic & Tetragonal & Cubic   \\  
   \hline          
Energy (meV/unit cell)  &  -39.31  &  -37.23  & -29.47 & 0  \\
  \hline
  \hline
 \end{tabular}
 \caption{Relative energies (potential energies) of different phases from DFT calculations. The cubic unit cell is chosen as the reference structure.}\label{t5}
\end{table}

\begin{table}
 \begin{tabular}{ P{3.5cm} | P{1.2cm} P{1.2cm} P{1.2cm} | P{1.2cm} P{1.2cm} P{1.2cm} | P{1.2cm} P{1.2cm} P{1.2cm} }
  \hline
  \hline
  &    \multicolumn{3}{c|}{R to O}         &     \multicolumn{3}{c|}{O to T}       &  \multicolumn{3}{c}{T to C}      \\
   \hline
Component  & $d_x$  & $d_y$ & $d_z$  & $d_x$   &  $d_y$  &  $d_z$  &  $d_x$  &  $d_y$ & $d_z$ \\
  \hline
Hardness--changing  & N  & N & Y  & Y   &  Y  &  Y  &  Y  &  Y & Y \\   
Displacive  & Y  & Y & Y  & Y   &  Y  &  N  &  Y  &  N & N \\ 
\hline
\hline
 \end{tabular}
 \caption{Phase--transition characters of each component. `Hardness--changing' includes bond softening and bond hardening, 
which are characterized by the change of the standard deviation of the Ti displacement distribution.
}\label{t6}
\end{table}

\newpage
\newpage
\clearpage

%\begin{figure}[htbp]
%\includegraphics[width=10.0cm]{angle.eps}
%\caption{The schematic figure of the angle potential in the modified bond valence model.}\label{f0}
%\end{figure}

\end{document}